\documentclass[%
 aip,
 amsmath,amssymb,
 reprint,%
]{revtex4-1}

\usepackage{graphicx}
\usepackage{dcolumn}
\usepackage{bm}

\usepackage[utf8]{inputenc}
\usepackage[T1]{fontenc}
\usepackage{mathptmx}
\usepackage{etoolbox}
\usepackage{optidef}

\makeatletter
\def\@email#1#2{%
 \endgroup
 \patchcmd{\titleblock@produce}
  {\frontmatter@RRAPformat}
  {\frontmatter@RRAPformat{\produce@RRAP{*#1\href{mailto:#2}{#2}}}\frontmatter@RRAPformat}
  {}{}
}%
\makeatother
\usepackage{hyperref}
\usepackage{comment} 
\usepackage{orcidlink}
\usepackage[normalem]{ulem}

\begin{document}

\title[BEC sub-wavelength confinement via superoscillations]{Bose-Einstein condensate sub-wavelength confinement via superoscillations}

\author{Dusty R. Lindberg\orcidlink{0001-5335-7941}}
\email{dustindberg@gmail.com}
\affiliation{Department of Physics and Engineering Physics, Tulane University, New Orleans, Louisiana 70118, United States}

\author{Gerard McCaul\orcidlink{0000-0001-7972-456X}}
\email{gmccaul@tulane.edu}
\affiliation{Department of Physics and Engineering Physics, Tulane University, New Orleans, Louisiana 70118, United States}

\author{Peisong Peng\orcidlink{0000-0002-7892-7231}}
\email{ppeng@tulane.edu}
\affiliation{Department of Physics and Engineering Physics, Tulane University, New Orleans, Louisiana 70118, United States}

\author{Lev Kaplan\orcidlink{0000-0002-7256-3203}}
\email{lkaplan@tulane.edu}
\affiliation{Department of Physics and Engineering Physics, Tulane University, New Orleans, Louisiana 70118, United States}

\author{Diyar Talbayev\orcidlink{0000-0003-3537-1656}}
\email{dtalbaye@tulane.edu}
\affiliation{Department of Physics and Engineering Physics, Tulane University, New Orleans, Louisiana 70118, United States}

\author{Denys I. Bondar\orcidlink{0000-0002-3626-4804}}
\email{dbondar@tulane.edu}
\affiliation{Department of Physics and Engineering Physics, Tulane University, New Orleans, Louisiana 70118, United States}

\date{\today}

\begin{abstract}
Optical lattices are essential tools in ultra-cold atomic physics. Here we demonstrate theoretically that sub-wavelength confinement can be achieved in these lattices through \textit{superoscillations}. This generic wave phenomenon occurs when a local region of the wave oscillates  faster than any of the frequencies in its global Fourier decomposition. To illustrate how sub-wavelength confinement can be achieved via superoscillations, we consider a one-dimensional tri-chromatic optical potential confining a spinless Bose-Einstein Condensate of $^{87}$Rb atoms. By numerical optimization of the relative phases and amplitudes of the optical trap's frequency components, it is possible to generate superoscillatory spatial regions. Such regions contain multiple density peaks at sub-wavelength spacing. This work establishes superoscillations as a viable route to sub-wavelength BEC confinement in blue-detuned optical lattices.
\end{abstract}

\maketitle

\section{Introduction}

Optical lattices are essential tools. They not only enable the controllable realisation of solid state models (including Josephson Junctions~\cite{Albiez2005, levy2007ac} and Hubbard models ~\cite{Mazzarella2006, lewenstein_ultracold_2007, heiselberg2009}), but also find application more broadly as (for example) optical tweezers~\cite{ashkin1970, moffitt2008recent, pesce2020optical}. Optical lattices are ubiquitous in ultra-cold atomic physics, especially in the confinement of Bose-Einstein Condensates~(BECs). Research into this exotic state of matter has grown tremendously since its first experimental observation \cite{letokhov_cooling_1976, anderson1995, davis1995}, and a key area of interest is sub-wavelength confinement within optical lattices~\cite{yi_state-dependent_2008, bienias_coherent_2020, nascimbene_dynamic_2015, jendrzejewski_subwavelength-width_2016, tsui-stroboscopic_subwavelength}. In this technique, ultra-cold atomic clouds are confined to distances smaller than the optical period (half wavelength, $\lambda/2$) of the fields that create the potentials.

One possibility for achieving sub-wavelength confinement is to exploit the phenomenon of \textit{superoscillations}. This corresponds to a field locally - i.e. within a limited spatial or temporal window - oscillating at a frequency exceeding its fastest Fourier component~\cite{berry_superoscillations_2018, BerryFourier, Rogers2020, chen_superoscillation_2019, jordan2025superoscillations}. This counterintuitive effect -- where local behavior is independent of global properties -- has applications in a broad range of research, including optical metrology~\cite{optical_ruler}, super-resolution microscopy~\cite{Lindberg_2012, Berry_2006}, super-transmission~\cite{Eliezer:14,zarkovsky_transmission_2020}, free-space plasmonics~\cite{Yuan2019}, 
quantum control~\cite{kempf_driving_2017, aharonov_how_2020},
radar~\cite{howell_super_2023, jordan_fundamental_2023}, and spectroscopy~\cite{mccaul_superoscillations_2023, peng_super-sensing_2025}.

Superoscillations however suffer from a significant drawback. The relative intensity of the field inside the superoscillating window is greatly suppressed relative to the field maxima. This is however less problematic in the ultra-cold atomic context, as low-amplitude, rapidly oscillating regions are sufficient to trap atoms. In this case the higher amplitudes outside a superoscillating region simply act as hard barriers to the atomic cloud. 

Superoscillations have been conjectured to offer a route for sub-wavelength confinement of BECs~\cite{Zhu2023_review}, but to date they have only been applied experimentally in the context of single-atom tweezers ~\cite{rivy_single_2023}. In this work, we theoretically demonstrate that sub-wavelength optical confinement of BECs can be achieved via superoscillations. Using an optical potential created via induced dipole force from off-atomic transition resonance (with two far-from-resonance frequencies), we simulate the sub-wavelength confinement of a D2 atomic transition line-resonant BEC of several thousand $^{87}$Rb atoms. Using off-resonant frequencies in optical potentials has a number of benefits, reducing both heating and spontaneous emission, as well as extending atom lifetime~\cite{Potvliege_2006, ADAMS19971, PhysRevA.47.R4567, GRIMM200095}. We first show that superoscillations can produce sub-wavelength spacing between isolated BEC peaks. We further demonstrate that the full condensate can be contained such that all density peaks remain below the diffraction limit, even if the condensate is not fully confined to the superoscillating region.

The rest of the paper is organized as follows: In Sec.~\ref{Sec2}, we describe the unique form a multi-wavelength optical lattice takes. In Sec.~\ref{Sec3}, the method is shown for obtaining superoscillations in a generalized, tri-chromatic optical potential. Finally, we utilize our unique solution to reduce the tri-chromatic potential to a bi-chromatic one and computationally show sub-wavelength optical confinement of a BEC cloud in a bi-chromatic superoscillating well.

\section{Optical Confinement of Bose-Einstein Condensates}\label{Sec2}

We consider the case of a BEC composed of $N=2000$ $^{87}$Rb atoms described by the Hamiltonian:
\begin{align}
    H =& \sum_{i=1}^{N}\left( -\frac{\hbar^{2}}{2m} \frac{\partial^{2}}{\partial\boldsymbol{r}_{i}^{2}}  + U\left( \boldsymbol{r}_{i} \right)\right) \nonumber \\
    &+ \sum_{i<j}\frac{4\pi\hbar^{2}a_{s}}{m}\delta \left(\boldsymbol{r}_{i} - \boldsymbol{r}_{j}\right). \label{eq:Hamiltonian_single}
\end{align}
Here, $m=1.443160 \times10^{-25}$ kg is the mass and $a_s =100a_0$ is the background s-wave scattering length of a single $^{87}Rb$ atom, where $a_0$ denotes the Bohr radius. 

Our goal is to find the form of the potential, $U(\boldsymbol{r})$, created by an external laser field. Since a $^{87}$Rb atom is electrically neutral and has no permanent dipole moment, the potential $U(\boldsymbol{r})$ is realized by inducing a dipole moment via an external laser field whose frequency, $\omega_{0}$, is resonant with one of the atomic transitions. The most commonly used transitions are D1 $\left(5^{2}\text{S}_{1/2}\rightarrow5^{2}\text{P}_{1/2}\right)$ and D2 $\left(5^{2}\text{S}_{1/2}\rightarrow5^{2}\text{P}_{3/2}\right)$ lines. 

Consider the case of two counter-propagating planar fields $\boldsymbol{E}_{\rm left}$ and $\boldsymbol{E}_{\rm right}$, such that the total electric and magnetic fields are
\begin{align}
    \boldsymbol{E} &= \boldsymbol{E}_{\rm left} + \boldsymbol{E}_{\rm right}, \label{eq:lr_Efield} \\
    \boldsymbol{B} &= \boldsymbol{B}_{\rm left} + \boldsymbol{B}_{\rm right} \nonumber \\
    &= \frac{1}{c}\hat{\boldsymbol{k}} \times \boldsymbol{E}_{\rm left} + \frac{1}{c}(-\hat{\boldsymbol{k}}) \times \boldsymbol{E}_{\rm right},    \label{eq:lr_Bfield}
\end{align}
where $\hat{\boldsymbol{k}}$ is the direction of propagation for the plane wave solution (see, e.g., Ch.~3~of~Ref.~\cite{milton2024classical}).  
If the counter-propagating fields are identical, $\boldsymbol{E}_{\rm left}(\boldsymbol{r}) = \boldsymbol{E}_{\rm right}(\boldsymbol{r})$, then it immediately follows that $\boldsymbol{B}=0$ identically, and therefore, 
\begin{align}
    -\frac{1}{c}\frac{\partial \boldsymbol{B}}{\partial t} = \nabla \times \boldsymbol{E} = \boldsymbol{0}. \label{eq:cp_assumption}
\end{align}
This is the typical setup used in an optical trap. The force exerted by an external electric field on an electric dipole $\boldsymbol{p}$ in the absence of a magnetic field [see, e.g., Eq.~(2.2)~of~Ref.~\cite{gordon_radiation_1973} or Sec.~4.1~of~Ref.~\cite{milton2024classical}] is
\begin{align}
    \boldsymbol{F} = \left( \boldsymbol{p} \cdot \nabla \right) \boldsymbol{E}.
\end{align}
Using the identity 
\begin{align}
    \nabla \left( \boldsymbol{p} \cdot \boldsymbol{E} \right) =& (\boldsymbol{p} \cdot \nabla) \boldsymbol{E} + (\boldsymbol{E} \cdot \nabla) \boldsymbol{p} \notag\\
    & + \boldsymbol{E} \times (\nabla \times \boldsymbol{p}) + \boldsymbol{p} \times (\nabla \times \boldsymbol{E}) \label{eq:dp_rule} 
\end{align}
as well as Eq.~\eqref{eq:cp_assumption} and the fact that the dynamical polarizability $\alpha$ does not have spatial dependence, the force on an induced dipole $\boldsymbol{p}= \alpha \boldsymbol{E}$ in the presence of a single external field [Eq.~(2.4)~of~Ref.~\cite{gordon_radiation_1973}, Eq.~(1.59)~of~Ref.~\cite{Steck_Teaching}] reads
\begin{align}
    \boldsymbol{F} =  \frac{1}{2}\nabla \left( \boldsymbol{p} \cdot \boldsymbol{E} \right), \label{eq:force_induced}
\end{align}
resulting in the well-known potential
\begin{align}
    U = -\frac{\alpha}{2}\boldsymbol{E}^2 \label{eq:singlefreqV}
\end{align}
for the case of a monochromatic field $\boldsymbol{E}$. 

For fields $\{ \boldsymbol{E}_j \}_{j=1}^3$ of three frequencies $\{ \omega_j \}_{j=1}^3$ separately satisfying Eqs.~\eqref{eq:lr_Efield} and \eqref{eq:lr_Bfield}, the total field and induced dipole moment become
\begin{align}
    \boldsymbol{E} 
    &= \boldsymbol{E}_1 + \boldsymbol{E}_2 + \boldsymbol{E}_3, \label{eq:induced_E_3field}\\
    \boldsymbol{p} &= \alpha_1 \boldsymbol{E}_1 + \alpha_2 \boldsymbol{E}_2 + \alpha_{3} \boldsymbol{E}_3, \label{eq:induced_dm_3field} 
\end{align}
where $\alpha_j$ are the dynamical polarizabilities
\begin{align}\label{eq:alphadef}
    \alpha_{j} = \frac{e^{2}}{m} \frac{1}{\omega_{0}^2 - \omega_{j}^2}.
\end{align}

Expanding Eq.~\eqref{eq:force_induced} with Eqs.~\eqref{eq:induced_E_3field} and \eqref{eq:induced_dm_3field}, we arrive at the sought force
\begin{align}
    \boldsymbol{F} = \frac{1}{2} \nabla[&\alpha_{1} \boldsymbol{E}_{1}^{2}  + \alpha_{2} \boldsymbol{E}_{2}^{2} + \alpha_{3} \boldsymbol{E}_{3}^{2} + \left( \alpha_{1} + \alpha_{2} \right) \boldsymbol{E}_{1} \boldsymbol{E}_{2} \nonumber \\ 
    &+ \left( \alpha_{1} + \alpha_{3} \right) \boldsymbol{E}_{1} \boldsymbol{E}_{3} +\left( \alpha_{2} + \alpha_{3} \right) \boldsymbol{E}_{2} \boldsymbol{E}_{3}] \label{eq:force_2field}
\end{align}
and the corresponding interaction potential
\begin{align}
    U = -\frac{1}{2}[&\alpha_{1} \boldsymbol{E}_{1}^{2}  + \alpha_{2} \boldsymbol{E}_{2}^{2} + \alpha_{3} \boldsymbol{E}_{3}^{2} + \left( \alpha_{1} + \alpha_{2} \right) \boldsymbol{E}_{1} \boldsymbol{E}_{2} \nonumber \\ 
    &+ \left( \alpha_{1} + \alpha_{3} \right) \boldsymbol{E}_{1} \boldsymbol{E}_{3} +\left( \alpha_{2} + \alpha_{3} \right) \boldsymbol{E}_{2} \boldsymbol{E}_{3}] \label{eq:dpV_3field} \,.
\end{align}

Instead of working with the many-body Hamiltonian~\eqref{eq:Hamiltonian_single}, we utilize the mean-field theory given by the time-independent Gross-Pitaevskii Equation (GPE)~\cite{pethick2008bose}
\begin{align} \label{eq:3DTimeIndependentGPE}
    \mu \Psi(\boldsymbol{r}) = 
    \left( -\frac{\hbar^2}{2m}\nabla^2+U(\boldsymbol{r})+g\left|\Psi(\boldsymbol{r})\right|^2 \right) \Psi(\boldsymbol{r}),
\end{align}
where $\Psi(\boldsymbol{r})$ is the ground state wave function normalized to one, $\int |\Psi(\boldsymbol{r})|^2 d^3 \boldsymbol{r} = 1$, $\mu$ is the chemical potential, and 
\begin{align}
    g = 4 \pi N \hbar^2 a_s /m, \label{eq:inter}
\end{align}
allowing total particle number $N$ to be changed easily by scaling integer multiples of $g$. We note quickly that the more common normalization for the GPE, $\int|\phi(\boldsymbol{r})|^2 d^3 \boldsymbol{r} = N$, is recovered by multiplying both sides of Eq.~\eqref{eq:3DTimeIndependentGPE} by $\sqrt{N}$ (as the number of particles is held as constant) and renormalizing the state to $\phi(\boldsymbol{r})=\sqrt{N}\Psi(\boldsymbol{r})$, such that the $N$-dependence found in $g$ in eq.~\eqref{eq:inter} is absorbed into $N|\Psi(\boldsymbol{r})|^2 = |\phi(\boldsymbol{r})|^2$.

\section{Speeding Oscillations by Minimizing Potential}\label{Sec3}

Setting the direction of propagation for our full field along the $x$-axis, such that $\boldsymbol{E}_j = (E_j(x), 0, 0)^T$, we can reduce the 3D geometry to 1D by tight harmonic confinement in the transverse $yz$ plane. Let us select confinement frequencies of 1000$\pi$ Hz for the $y$ and $z$ axes, and a weak harmonic confinement of 0.5$\pi$ Hz for the $x$-axis. This creates a cigar-shaped potential that   may confine the BEC to 1D without interfering with the superoscillating potential created inside, as the effect of the confining $x$-axis $0.5\pi$ Hz harmonic is negligible at distance scales on the order of $\sim 100\mu$m. Hence, Eq.~\eqref{eq:3DTimeIndependentGPE} is replaced by the following 1D GPE  (see, e.g.,  Refs.~\cite{Lindberg_2023, Boegel_2021})
\begin{align} \label{eq:1DTimeIndependentGPE}
    \mu\psi(x) =& \left( -\frac{\hbar^2}{2m}\frac{\partial^2}{\partial x^2}+V(x)+ g\left|\psi(x)\right|^2 \right) \psi(x),
\end{align}
where the ground state $\psi(x)$ is normalized as $\int  |\psi(x)|^2 dx = 1$. 

For a monochromatic field producing the trapping potential~\eqref{eq:singlefreqV}, the minimum confinement distance achievable is dictated by the diffraction limit, which is simply half the wavelength, $\lambda/2$, of the laser field. 

Consider now the three-color field from Eq.~\eqref{eq:induced_E_3field} defined by
\begin{align}
    E(x) &= \sum_{j=1}^{3} E_j(x), \label{eq:control_field}\\
    E_j(x) &= A_{j}\cos\left(k_{j}x + \phi_{j}\right) \notag  \\
    &=  b_{j}\sin\left(k_{j}x\right) + c_{j}\cos\left(k_{j}x\right), \label{eq:field_mod}
\end{align}
where $k_j = 2\pi/\lambda_{j}$, $\lambda_1=702$ nm, $\lambda_{2} = 728$ nm, and $\lambda_{3} = 780$ nm. By adjusting the amplitudes $A_j$ and phases $\phi_j$ (or equivalently $b_j$ and $c_j$, see Eq.~(4.21.1\_5) of Ref.~\cite{NIST:DLMF}), we can destructively interfere the three fields to achieve superoscillations within a spatial interval 
\begin{align}
    x_{\rm super} = [-125, \, 125] \text{ nm}. \label{eq:so_region}
\end{align}
Note that the width of $x_{\rm super}$ is  $250$ nm, which is 0.356$\lambda_1$. The resulting potential will have a full period width of the least common multiple of the individual wavelengths, lcm(702, 728, 780) nm = 98.280 $\mu$m. As such, a range that contains the full period of the complete standing wave is
\begin{align}
    x_{T} = [-49.140, \, 49.140] \mu{\rm m}. \label{eq:pot_width}
\end{align}

Superoscillations of a laser field have been synthesized via intensity minimization~\cite{mccaul_superoscillations_2023}. We extend this idea to minimizing $\int_{x_{\rm super}} Vdx$ with respect to $b_j$ and $c_j$, where the potential $V$ is induced by the trichromatic field~\eqref{eq:control_field}. To enforce physical units and avoid the trivial solution, we add the constraint $\int_{x_{T}} V dx = V_0$ for some constant $V_0$. 

Hence, the values of the parameters $\{b_j, c_j\}_{j=1}^3$ that are candidates for superoscillations are obtained by 
\begin{mini}[s]
    {\{b_j, c_j\}_{j=1}^3}{\int_{x_{\rm super}}\widetilde{V} dx}{}{}
    \addConstraint{\int_{x_{T}} \widetilde{V} dx = 1}, 
    \label{eq:minFormulation} 
\end{mini}
with
\begin{align}\label{eq:TildeV}
    \widetilde{V} = \frac{V}{V_0} = -&\frac{1}{2 V_0}\big[\alpha_{1} E_{1}^{2}  + \alpha_{2} E_{2}^{2} + \alpha_{3} E_{3}^{2} + \left( \alpha_{1} + \alpha_{2} \right) E_{1} E_{2} \nonumber \\ 
    &+ \left( \alpha_{1} + \alpha_{3} \right) E_{1} E_{3} +\left( \alpha_{2} + \alpha_{3} \right) E_{2} E_{3}\big].
\end{align}
The form of $\widetilde{V}$ follows from  Eq.~\eqref{eq:dpV_3field}, and
the dynamical polarizabilities $\alpha_j$ are defined in Eq.~\eqref{eq:alphadef}, where $\omega_0=2\pi \cdot384.230$~THz, corresponding to the widely used D2 transition line ($\lambda_0 = 780.24$ nm).
The scaling constant $V_0$ is chosen to make the objective function and constraint unitless and optimization numerically well conditioned, i.e., by avoiding very small or large numerical values.

The problem~\eqref{eq:minFormulation} can be readily solved by the method of Lagrange multipliers. First, we rewrite the objective function as a quadratic form 
\begin{align}
    \int_{x_{\rm super}} \widetilde{V}dx = \boldsymbol{a}^{\rm T} S(x_{\rm super}) \boldsymbol{a}, 
\end{align}
where $\boldsymbol{a}^{\rm T} = (b_1, c_1, b_2, c_2, b_3, c_3)$ is the vector of unknowns and $S(x_{\rm super})$ is a $6\times 6$  symmetric matrix. The constrained minimization~\eqref{eq:minFormulation} is reduced to finding the extrema of
\begin{align}
    \mathcal{L} = \boldsymbol{a}^{\rm T}&S(x_{\rm super})\boldsymbol{a} 
    -\lambda  \left(\boldsymbol{a}^{\rm T}S(x_{T})\boldsymbol{a} -1\right).
\end{align}
The sought extrema are eigenvectors of the following generalized eigenvalue problem
\begin{align}
    S(x_{\rm super})\boldsymbol{a} = \lambda S(x_{T}) \boldsymbol{a}, \label{eq:eignSolution}\\
    \boldsymbol{a}^{\rm T}S(x_{T}) \boldsymbol{a} = \mathbb{I}.
\end{align}

\begin{figure*}
    \centering
    \includegraphics[width=0.9\linewidth]{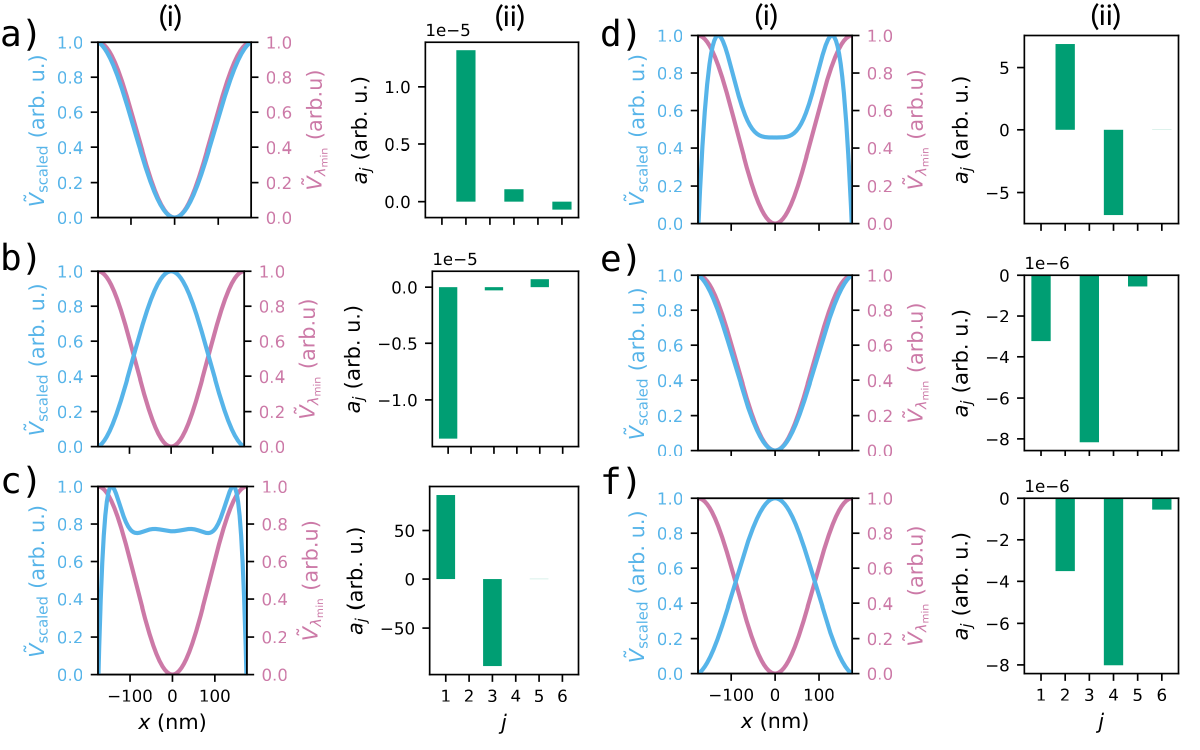}
    \caption{The six solutions \textbf{(a - f)} to the three-frequency eigenvalue equation shown over the scale of a single period of the shortest wavelength, $\lambda_{min}/2 = 351$ nm, centered about 0. Column \textbf{(i)} demonstrates the potential created by each solution. For illustrative purposes only, we have scaled each potential $\tilde{V}$ such that, within the superoscillating region, the amplitude ranges from 0 to 1. Column \textbf{(ii)} shows the coefficients required to make $\tilde{V}_{\rm scaled}$. Note that the solutions favor exclusively either the sine or cosine amplitudes.
    We can see in (i) that solutions (a) and (e) basically overlap with the shortest wavelength field $\lambda_{min}=702$ nm, and that (b) and (f) are just a phase shift from (a) and (e). However, in (c) and (d), we see definite evidence of superoscillations both in the potential shape and in the correspondingly large coefficients required to reach the same amplitude as in (a-b) and (e-f).}
    \label{fig:BEC_eigen}
\end{figure*}

Solving the eigenvalue problem~\eqref{eq:eignSolution}, we obtain six eigenvectors that provide values of the parameters for the potential~\eqref{eq:TildeV}, as shown in Fig.~\ref{fig:BEC_eigen}. Of these six candidates, the potentials shown in Figs.~\ref{fig:BEC_eigen}(c) and \ref{fig:BEC_eigen}(d) exhibit superoscillations. We can also see that the coefficients of the sinusoidal fields needed to achieve a maximum amplitude of 1 within $x_{\rm super}$ are several orders of magnitude higher than outside of the non-superoscillating solutions. This is one of the major trade-offs in working with superoscillations, as the oscillations within the window result from destructive interference and are accordingly very low. We choose to proceed with the solution from Fig.~\ref{fig:BEC_eigen}(d), as the required coefficients are smaller than in Fig.~\ref{fig:BEC_eigen}(c) and the deeper central well with shallow wells to the left and right allow for the possibility of three density peaks of sub-wavelength spacing.

\begin{figure}
    \centering
    \includegraphics[width=0.9\linewidth]{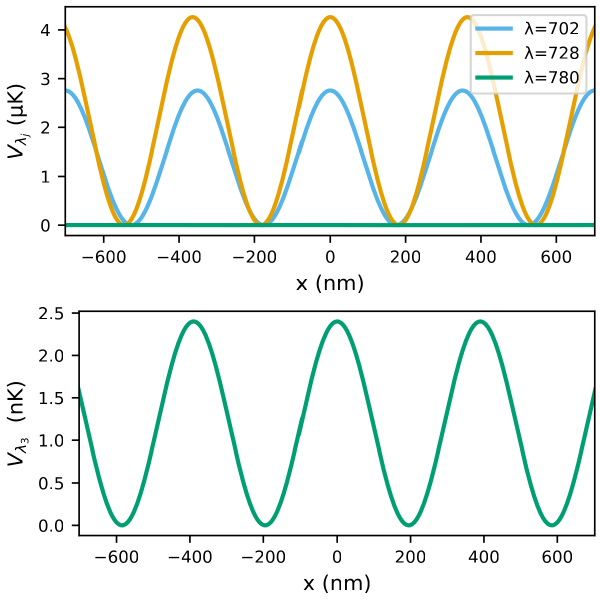}
    \caption{A demonstration of the potential created by each individual wavelength field, $V_{\lambda_j}$. We note that the magnitudes of the two far-off-resonance wavelengths are dominant, on the order of $\mu$K, while we must actually move to the nK scale to see the near-resonance wavelength field.}
    \label{fig:indv_fields}
\end{figure}

\begin{figure}
    \centering
    \includegraphics[width=0.75\linewidth]{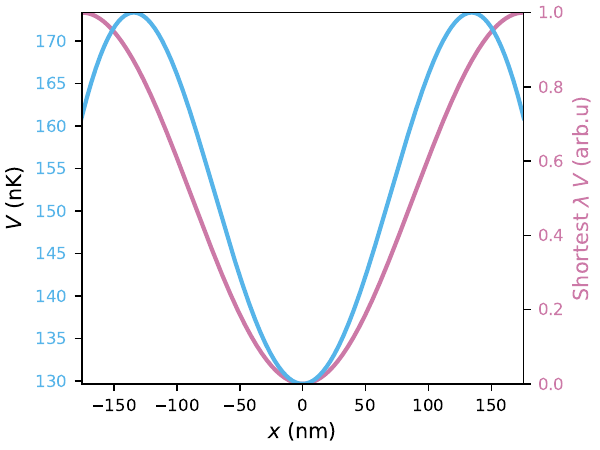}
    \caption{The bi-chromatic solution given by using only the $E_1$ and $E_2$ fields. We see that the potential is still superoscillating, but with a smoother central well. We also see that the potential does not decrease as fast as with the tri-chromatic solution at the edges of our chosen window, showing that there is some loss due to the exclusion of the third, near-resonance field.}
    \label{fig:2field_solution}
\end{figure}

Using the solution from Fig.~\ref{fig:BEC_eigen}(d) to simulate a specific case of a BEC, we will choose $V_{0}$ such that the normalized potential is scaled up to a maximum of $14\mu$K. While this scaling can potentially push the amplitudes into difficult-to-achieve ranges, this is the tradeoff for achieving sub-wavelength confinement using superoscillations. For comparison, we depict in Fig.~\ref{fig:indv_fields} the potentials induced by the individual fields 
\begin{align}
    V_{\lambda_j} = -\frac{\alpha_j}{2}E_j^2.
\end{align}
Looking at Fig.~\ref{fig:indv_fields}, we can see the physical maximum of each individual wavelength potential, with the two far-off-resonance fields reaching maxima of $V_{\lambda_1}\lessapprox4\mu$K and $V_{\lambda_2}\lessapprox3\mu$K respectively. Quite beneficially, the near-resonance wavelength potential is three  orders of magnitude lower than the far-off-resonance magnitudes, reaching a maximum of only $V_{\lambda_3}\lessapprox2.5$nK. While this is somewhat high overall, the maximum within the range of $x_{\rm super}$ is $\approx 173.4$ nK, allowing the low-energy BEC to concentrate in wells of our superoscillating region. Finally, we set the region over which we will be simulating the BEC. Because the amplitude for $E_3$ is so low, we investigate the possibility of confinement with only the two far-off-resonance fields, setting $E_3$ to 0. We can see from Fig.~\ref{fig:2field_solution} that the resulting bi-chromatic potential is still superoscillating, even when removing the third wavelength, although the shape has changed to be far smoother than the flat central well of Fig.~\ref{fig:BEC_eigen}(d). We also note the decreased oscillation strength of the bi-chromatic potential at the edges of $x_{\rm super}$ compared to the tri-chromatic solution. While this does represent a trade-off, it is more than welcome given the benefits of using only far-off-resonance wavelengths and decreasing experimental complexity.

\begin{figure*}[h]
    \centering
    \includegraphics[width=0.75\linewidth]{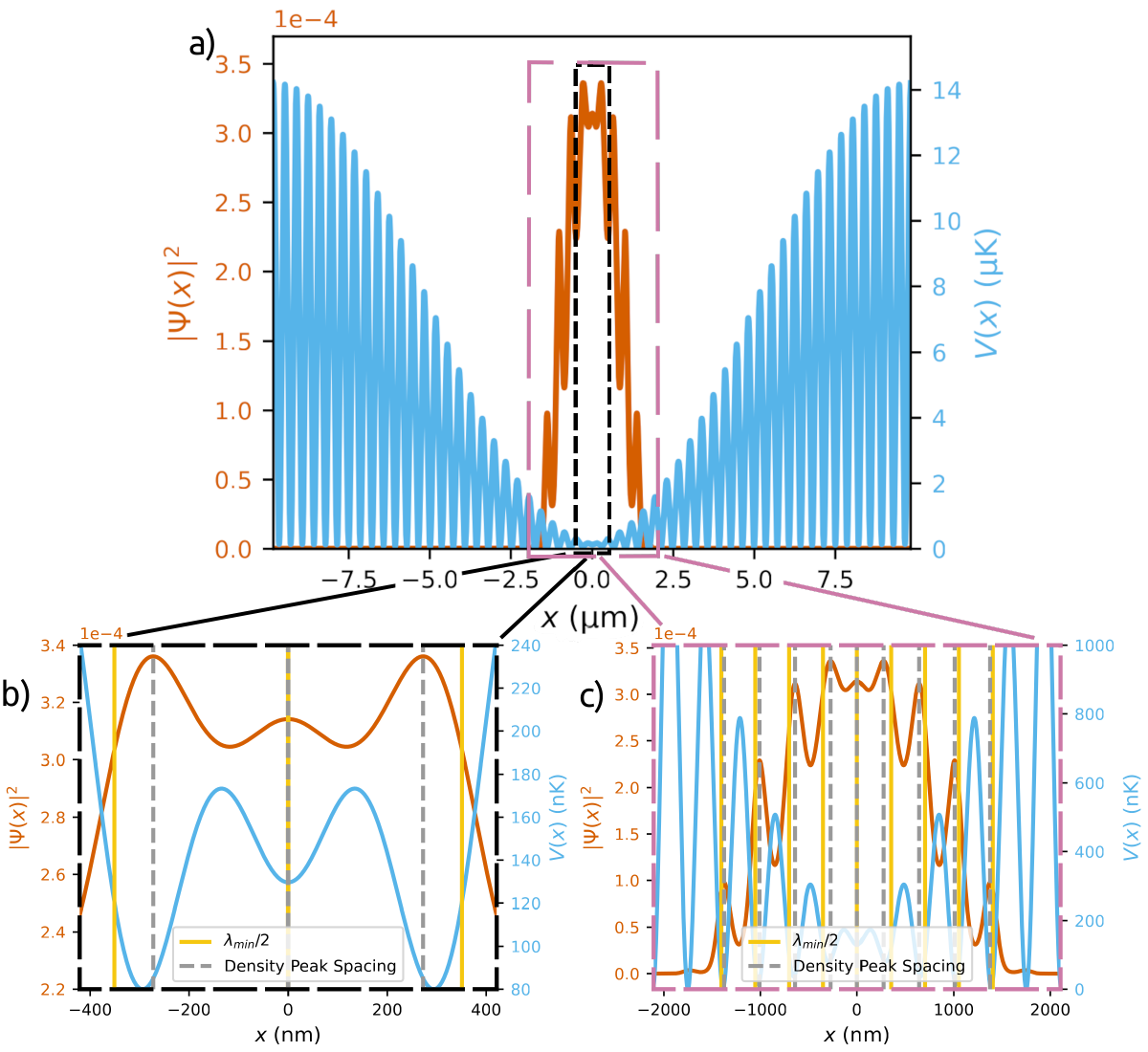}
    \caption{\textbf{a)} The BEC density within a single superoscillating well. We include two magnified regions: \textbf{b)} The black enclosed region shows the optimal sub-wavelength spacing of BEC peaks. \textbf{c)} The purple enclosed region shows the entire density profile. Note that even though the peak spacings increase as one moves further from the superoscillating region, all nine peaks are contained in a region smaller than 9$T_{\rm opt}$.}
    \label{fig:BEC_groundstate}
\end{figure*}

We then use the imaginary time method to find the ground state  of the BEC [Eq.~\eqref{eq:1DTimeIndependentGPE}] in the superoscillating potential seen in Fig.~\ref{fig:BEC_groundstate}, resulting in the ground state energy $\approx 303.8$ nK. Looking at Fig.~\ref{fig:BEC_groundstate}, we see that the BEC occupies the central region, forming nine clearly defined peaks within a region of $\approx 1.5\mu$m. With an optical lattice constructed using our shortest wavelength, $\lambda_1 = 702$ nm, the spacing between density peaks cannot be reduced below the optical period length of the shortest pulse, $T_{\rm opt} = \lambda_1/2 = 351$~nm, which is equivalent to the diffraction limit. By zooming into the black outlined region in Fig.~\ref{fig:BEC_groundstate}, we can see that the three most prominent peaks are spaced below $T_{\rm opt}$, with the exact spacing being 0.7759$T_{\rm opt}$, or $\approx 272.3$ nm. Zooming out to the purple outlined region of Fig.~\ref{fig:BEC_groundstate} to show the full density profile, we can see that the peak spacings increase symmetrically on either side. The next spacings are 1.0493$T_{\rm opt}$, 1.0493$T_{\rm opt}$, and 1.0459$T_{\rm opt}$ as we move away from the center. Notably, though, the expanded spacings only very slightly exceed $T_{\rm opt}$. For nine evenly spaced wells of wavelength $\lambda_1$, we would expect the distance between the outermost peaks to be 8$T_{\rm opt}$, or 2808 nm. However, the total distance between our outermost peaks is $\approx 2752.1$ nm, or 7.8408$T_{\rm opt}$.

\section{Conclusion}

We demonstrated sub-wavelength confinement of a BEC using superoscillations in optical lattices. Our approach employs a minimization method that creates superoscillations through destructive interference between fields of different wavelength. 

We used two far-off-resonance wavelengths, ($\lambda_1 = 702$ nm and $\lambda_2 = 728$ nm),  to trap $^{87}$Rb atoms. This combination produced superoscillations within a 250 nm window, smaller than half the shortest wavelength. The resulting potential confined a BEC of $2000$ atoms with a ground state energy of  303.8 nK. The central density peaks achieved sub-wavelength spacing of 272.3 nm. This represents 0.776 times the optical period of the shortest wavelength component. Even accounting for the wider spacings between the outer peaks, the total BEC width remained below the diffraction limit. Nine density peaks fit within 7.84 optical periods, compared to the expected 8 periods for conventional confinement.

Our results confirm that superoscillations provide a viable path to sub-wavelength BEC confinement. The trade-off is the requirement for high field amplitudes outside the superoscillating region. This approach opens new possibilities for ultra-cold atom manipulation beyond traditional diffraction limits. Future work could explore optimization of field parameters to reduce required amplitudes while maintaining sub-wavelength confinement.

\acknowledgments

The authors are grateful to Prof.~Sergey N. Shevchenko and Prof.~Andrii G. Sotnikov for organizing the Kharkiv Quantum Seminar online~\footnote{\url{https://sites.google.com/view/kharkivquantumseminar}}. The idea of this paper was conceived during the seminar of Prof.~Gediminas Juzel\={u}nas.

At the initial stages, this work was supported by the W. M. Keck Foundation. D.R.L. is supported by NASA EPSCoR and/or the Board of Regents Support Fund. D.I.B. was supported by Army Research Office (ARO) (grant W911NF-23-1-0288; program manager Dr.~James Joseph). The views and conclusions contained in this document are those of the authors and should not be interpreted as representing the official policies, either expressed or implied, of ARO or the U.S. Government. The U.S. Government is authorized to reproduce and distribute reprints for Government purposes notwithstanding any copyright notation herein.

\section*{Code and data availability}

The numerical codes used in this study together with the raw data are available at~\footnote{\url{https://github.com/dustindberg/GPE/blob/master/BEC_ThreeFieldExample.ipynb}}.

\bibliography{sobec}

\end{document}